\documentclass[sigconf, screen]{acmart}

\usepackage{algorithm}
\usepackage{algpseudocode}
\usepackage{float}
\usepackage{afterpage}

\AtBeginDocument{%
  }

\copyrightyear{2024}
\acmYear{2024}
\setcopyright{acmlicensed}\acmConference[MIG '24]{The 16th ACM SIGGRAPH Conference on Motion, Interaction, and Games}{November 21--23, 2024}{Arlington, VA, USA}
\acmBooktitle{The 16th ACM SIGGRAPH Conference on Motion, Interaction, and Games (MIG '24), November 21--23, 2024, Arlington, VA, USA}
\acmDOI{10.1145/3677388.3696327}
\acmISBN{979-8-4007-1090-2/24/11}

\definecolor{myBlack}{rgb}{0,0.,0}
\definecolor{myRed}{rgb}{1,0.,0.}
\definecolor{myLightRed}{rgb}{0.7,0.,0.}
\definecolor{myOrange}{rgb}{0.8,0.3,0.}
\definecolor{myGreen}{rgb}{0,0.5,0.27}
\definecolor{myBlue}{rgb}{0,0.0,0.9}
\definecolor{myPurple}{rgb}{0.6,0.,0.6}
\definecolor{myBluegreen}{rgb}{0.2,0.6,0.8}

\newcommand{\yuchen}[1]{\textcolor{myBlack}{#1}}

\begin{document}

\title{Real-time Diverse Motion In-betweening with Space-time Control}

\author{Yuchen Chu}
\email{yuchen_chu@sfu.ca}
\affiliation{%
  \institution{Simon Fraser University}
  \country{Canada}
}

\author{Zeshi Yang}
\email{zs243@mail.ustc.edu.cn}
\affiliation{%
  \institution{Independent Researcher}
  \country{China}
}


% \author{}
% \affiliation{%
%   \institution{}
%   \city{}
%   \country{}}
% \email{}

\renewcommand{\shortauthors}{}

\begin{abstract}

In this work, we present a data-driven framework for generating diverse in-betweening motions for kinematic characters. Our approach injects dynamic conditions and explicit motion controls into the procedure of motion transitions. Notably, this integration enables a finer-grained spatial-temporal control by allowing users to impart additional conditions, such as duration, path, style, etc., into the in-betweening process. We demonstrate that our in-betweening approach can synthesize both locomotion and unstructured motions, enabling rich, versatile, and high-quality animation generation.
\end{abstract}

\begin{CCSXML}
<ccs2012>
   <concept>
       <concept_id>10010147.10010371.10010352</concept_id>
       <concept_desc>Computing methodologies~Animation</concept_desc>
       <concept_significance>500</concept_significance>
       </concept>
 </ccs2012>
\end{CCSXML}

\ccsdesc[500]{Computing methodologies~Animation}
\ccsdesc[300]{Computing methodologies~Motion synthesis}
\keywords{character animation, motion-synthesis, in-between, deep learning}

\begin{teaserfigure}
  \includegraphics[width=\textwidth]{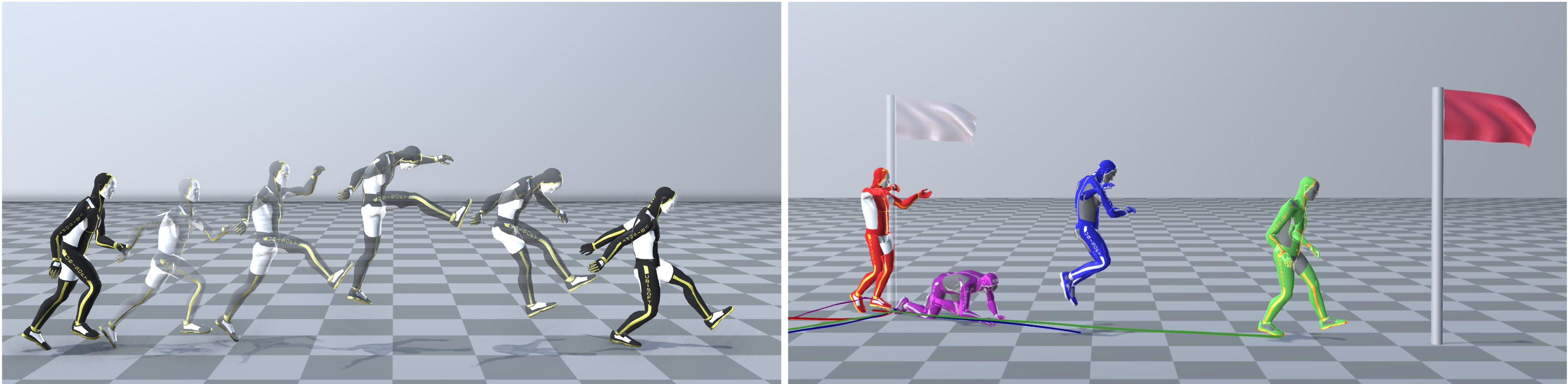}
  \caption{We present Dynamic Conditional Mixture-of-Experts(DC-MoE) neural network, a framework that enables motion transition for in-betweening, supporting diverse styles, durations, and trajectories. From left to right: a) a motion sequence generated based on the target key pose; b) diverse motions generated for the same target position.}
  \label{fig:teaser}
\end{teaserfigure}


\maketitle

\section{Introduction}
Given some sparse character poses, how can we synthesize in-between motions being both controllable and diverse? This problem is at the core of the interactive animation system, which can be applied in various applications including animation software, video games, and virtual reality. The ideal solution should be able to produce high-quality motion, be computationally efficient, and support a variety of controllable conditions. However, meeting all these requirements simultaneously is a challenging problem.

In this paper, we tackle this problem in two steps. First, we learn a kinematic pose prediction model of character motions using pairs of current and target frames. Motivated by the recent success of the Mixture-of-Experts (MoE) scheme for kinematic motion 
generation\cite{SCA2023:PhaseInbetween,sig23:tang_realtime_stylized,sig2022:soccerMoE,Sig2020:MotionVAE}, we introduce a Dynamic Conditional Mixture-of-Experts (DC-MoE) framework, which utilizes the MoE scheme in a dynamic conditional format for autoregressive in-betweening pose estimation. Given the current frame, target frame, and condition, the DC-MoE predicts the pose and temporal features at the next time step. Once the DC-MoE is learned, given the current and target key poses, it can generate diverse in-between motions by controlling motion attributes like style and duration. Next, we inject explicit control over the character's root trajectory to reach the target via trajectory guidance. Unlike many previous learning-based approaches, which estimate the character trajectory using neural networks in an implicit way, we show that by leveraging the principle of motion matching in our framework, we can synthesize a diverse range of root trajectories connecting the current pose and the target pose, covering various speeds, root positions, and durations. The effectiveness of our framework is evaluated on 100STYLE dataset\cite{2022:100StyleDataset} and LaFAN1 dataset\cite{Sig2020:RobustIbt} under multiple metrics in terms of motion quality.

In conclusion, our main contributions can be summarized as follows:
% itemize
\begin{itemize}
\item We introduce a motion synthesis framework DC-MoE, capable of generating in-betweening human motions with diverse styles under varied time durations.  
\item We propose a method to incorporate trajectory guidance with in-betweening motion, enabling controllable synthesis procedures.
% \item \textbf{\textit{{Method of balancing the style and target reaching precision. \textit{(Currently there's limitation and need for improvements)}}}}
\end{itemize}

\section{Related work}
\subsection{General Motion synthesis}
Incorporating interaction into the motion synthesis process is crucial for ensuring controllable motion generation. % can provide user-oriented flexibility and 
Graph-based methods \cite{Sig2002:MotionGraph,MIG2007:paramterizedmotiongraph,2008:motifgraph} enable motion synthesis via traversal through nodes and edges representing similar character states. 
As graph construction and distance map maintenance grow increasingly complex with large datasets, Motion Matching\cite{MotionMatching2016:GDC}, a greedy approximation of the Motion Field\cite{sig2010:motionField} using distance metrics to find the best matching frame was developed. Despite improving the scalability of animation systems, maintaining the whole motion database during runtime is highly memory intensive.

% paragraph: network -> deterministic model for improving naturalness
Leveraging neural networks to model motion representations has shown great advancements in scalability and computation efficiencies\cite{sig20:learnedmotionmatching, sig:MotionSynthesisandEditing, Sig2020:MotionVAE}.
One common challenge in using neural network predictions is maintaining the naturalness of the predicted character motions. %and considerable efforts have been dedicated to this area.\cite{sig2016:HoldenCNNFramework} use Convolutional Neural Network along the time domain to learn the motion manifold and then take control signal like root trajectory as input to synthesize constrained motions offline. Many auto-regressive deep learning methods that leverage previous frame information have been proposed. 
\cite{sig2017:PFNN} take the phase function to globally modify network weights, enabling cyclic motion generation and preventing convergence to an average pose. \cite{Sig2018:MANN_mix_Expert} propose to use neural gating functions to synthesize complicated quadruped locomotion. \cite{Sig2019:neuralStateMachine} further extend the phase value with gating network \cite{Sig2018:MANN_mix_Expert} to support environment interaction with different actions. However, a single phase can not guarantee the smooth transitions of different actions, which tends to cause discontinuity. To address this, local phases for contact-rich interaction\cite{sig2020:localPhase} and latent phases\cite{sig2022:deepPhase} for modeling whole body movement are successively proposed.

% paragraph: network -> generative model for improving diversity
Another common challenge for prediction-based motion generation is the lack of diversity in the synthesized motion. 
\cite{Sig2020:Moglow} use a normalizing-flows-based method to model the exact maximum likelihood of motion sequence. \cite{Sig2020:MotionVAE} use a Variational Auto-Encoder(VAE) to learn a latent motion space conditioned on the previous motion frame, supporting probabilistic motion sampling and thus enriching motion variations. \cite{sig22:Ganimator} applies generative adversarial networks (GAN) on a single training motion sequence to synthesize plausible variations.

Recently, diffusion-based methods\cite{ICLR2023:MDM,ICCV2023:guidedMDM,sig23:audioDrivenDiffusion, cvpr2023:EDGE,sig2024:flexibleIBT,raab2024monkeyseemonkeydo} have shown significant progress in diverse motion synthesis. \yuchen{\cite{ICLR2023:MDM, sig24:AMDM, ICLR2024:OmniControl} incorporate diffusion models with interactive control. In particular,  \cite{ICLR2023:MDM} and \cite{sig24:AMDM} demonstrate that either spatial or temporal in-painting for missing joints can be achieved through a pre-trained diffusion model, with in-betweening as a downstream task.} Although the model allows partial overwriting of the noisy input with conditional key-frames or joints, it does not ensure the generated motion can follow the condition and may yield physically implausible motions. The work of \cite{sig2024:flexibleIBT} present a flexible in-betweening pipeline through random sampling keyframes and concatenating binary masking during training. However, the prolonged reverse diffusion steps preclude their use in real-time interaction scenarios.

\begin{figure*}[!t] %
    \centering
    \includegraphics[width=\linewidth]{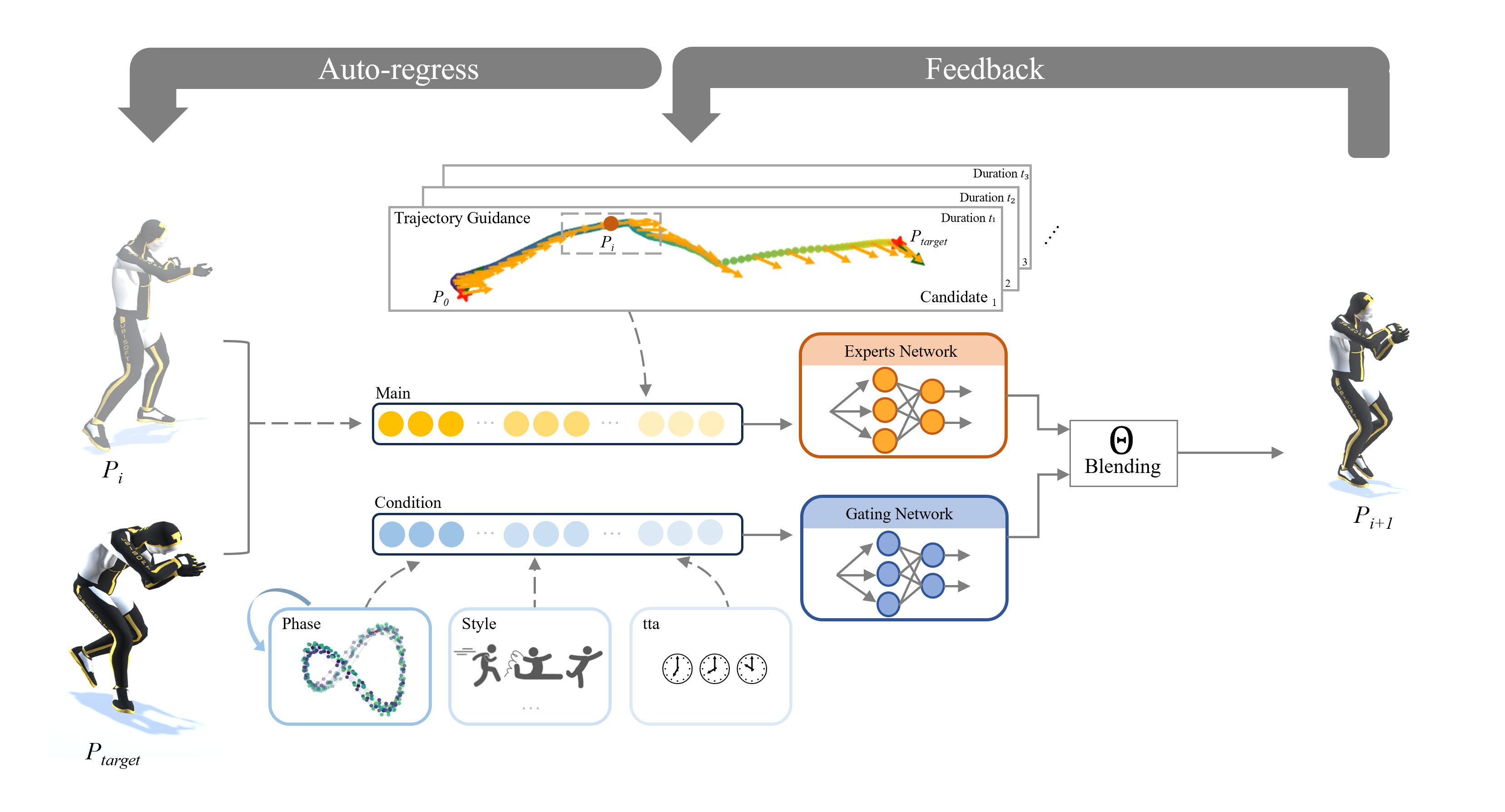} % Adjust width as needed
    \caption{The DC-MoE framework consists of two parts. The gating network takes the condition that contains the phases predicted by the network in an autoregressive manner, along with style and the time-to-arrive at the target pose as inputs, and outputs gating weights. The expert network takes the current pose state $P_{i}$, target pose state $P_{target}$, along with the root trajectory centered at  $P_{i}$  as inputs, and gets blended with the gating weights to output the pose features in the next timestamp.
    \label{fig:framework}}
\end{figure*}

\subsection{Motion In-betweening}
% {I should mention the difference of our method to strengthen more on the planning for the missing part.}
The fundamental objective of motion in-between is to predict reasonable motions while satisfying constraints set by both past and target future frames. While our discussion primarily centers on kinematics, it is noteworthy that physics-based methods such as space-time constraints based on Newtonian physics with keyframes, can also be considered as in-betweening techniques \cite{tog2002:Karenspacetime,tog2003:nancySpaceTime}.

The most straightforward kinematics solution in most commercial authoring tools involves spherical linear interpolation(SLERP) for rotation and spline interpolation for translation, such as Catmull-Rom\cite{catmull1974class}and Bezier curves. However, linear interpolation alone cannot produce realistic motions, demonstrating rigid and robot-like animations.

Neural network methods have been applied to the in-between task. \cite{2dJump:VandePan} utilize a Recurrent Neural Network(RNN) for generating 2D jump animation based on specific keyframes. 
\cite{sig2018:RTN} employ the Recurrent Transition Networks (RTN), built on the Long Short-Term Memory (LSTM) model, and this framework is further extended by \cite{Sig2020:RobustIbt} through the incorporation of scheduled target noise to generate varied motions over different in-between durations. However, it is noticeable that these variations are restrained, primarily due to the mode collapse issue of GAN, which often results in generating similar neutral motions.
\cite{Sig2022Asia:twoStangeTransformer} propose a transformer-based framework with a coarse-to-fine approach, using two-stage transformer encoders.  
While using multiple starting frames may enhance transition continuity for multiple in-between sections, the necessity to set many frames (e.g. 10 or more) might overburden the artists during the authoring process, thereby limiting the efficiency and overall practicality of this method.
 \cite{sig2022:cvae} incorporate LSTM with Conditional Variational Autoencoder(CVAE) to model a low-level motion transition and show fixed frames in-between motion generation in real-time. They further explore the integration of motion-inbetweening with style transfer by \yuchen{style encoding and learned phase manifold} in an online learning fashion\cite{sig23:tang_realtime_stylized}. 
 However, the prominence of the guided style within the generated transition largely depends on the setting of the time duration and target poses. 
 % Meanwhile, motions with subtle style differences show similar phase latent codes on the phase manifold, making \cite{sig23:tang_realtime_stylized} difficult to achieve fine-grained style control. 
 
 More recently, \cite{SCA2023:PhaseInbetween} integrate the autoregressive model with phase variables and trajectory to improve continuous transitions. This integration has proven useful when the target poses are not in the training data distribution. However, \cite{SCA2023:PhaseInbetween} needs to manually set the durations for all in-between trajectories, causing unstable outcomes when the set durations do not match the training distribution. Furthermore, predicting the next pose without given styles tends to generate averaged poses. In contrast, our method enhances motion in-betweening by incorporating both style control and trajectory guidance. This addition provides more controllability both in space and time.

\section{Methodology}
In this section, we describe the framework of DC-MoE that predicts the in-between motions from the input motion pairs. An overview of our framework is provided in Figure~\ref{fig:framework}. We first describe the network's structure, then the training process of the network.

\subsection{Network Structure}
Our framework is an autoregressive model that predicts $ \mathrm{Y}$, the state of the character in the next frame, given $ \mathrm{X}$, the state in the current frame and the target frame, incorporating conditional control signals $\mathrm{C}$. We utilize a Mixture-of-Expert architecture similar to \cite{Sig2018:MANN_mix_Expert,SCA2023:PhaseInbetween}. We additionally enforce the gating network to be informed of the motion style and time to reach the target. This allows the model to predict the motion with diversity instead of outputting averaged motions. 

\subsubsection{Expert Network}
The input features for the expert network consists of three components $ \mathrm{X_{main}}=\mathrm\{{X^{\mathcal{C}}}, X^{\mathcal{T}}, X^{\mathcal{R}}\}$. $\mathrm{X^{\mathcal{C}}}$ are the state of the character in the current frame which contains ${\{{p_i}, \dot{p}_i, {r_i}\}} \in \mathbb{R}^{12}$, where ${p_i} \in \mathbb{R}^{3}$ is the bone position, ${\dot p_i} \in \mathbb{R}^{3}$ is the bone linear velocity and ${r_i} \in \mathbb{R}^{6}$ is the bone rotation expressed in 6D vectors~\cite{Zhou_2019_CVPR}. Here $i\in \mathbb{J}$, where $\mathbb{J}=22$ denotes the number of joints. $\mathrm{X^{\mathcal{T}}}$ are the target state of the character containing $\mathrm\{{p_i}, {r_i}\}  \in \mathbb{R}^{6}$, with the same definition as in ${X^{\mathcal{C}}}$. And $ X^{\mathcal{R}}$ are the root trajectory projected on the ground which includes a set of $\mathrm\{{r_t}, \dot{r_t}, {d_t}\} \in \mathbb{R}^{6}$, where ${r_t}\in \mathbb{R}^{2}$ is the root position, $\dot{r_t}\in \mathbb{R}^{2}$ is the root velocity and ${d_t}\in \mathbb{R}^{2}$ is the root facing forward direction. We follow \cite{SCA2023:PhaseInbetween} to construct a $13$-frame window where $t$ ranges from $-6$ to $6$ with the current state in the middle, along with 6 average sampling time-stamps in both the past and future temporal domain. 

The expert network is a three-layer fully-connected network, and each layer has 512 hidden units followed by SiLU activation. We choose the expert number $K=16$, which outputs 16 sets of expert weights from the input $\mathrm{X_{main}}.$ We find that 16 experts are enough to generate natural stylized motions, and increasing the latent size will not further improve the quality of the generated motions.

\subsubsection{Gating Network} 
We define the conditioned feature for the gating network $ \mathrm{C_{cond}}=\mathrm\{{C^{\mathcal{P}}}, C^{\mathcal{S}}, C^{\mathcal{T}}\}$. $ \mathrm{C^{\mathcal{P}}}$ is the phase feature consisting of the set $ {\mathrm\{{(p_0^{n}}, {p_1^{n}),....,(p_{2i-1}^{n}}, {p_{2i}^{n})}\}}$, as proposed in DeepPhase~\cite{sig2022:deepPhase}. Here $n \in \mathbb{R}^{10}$ is the phase channels and $i \in \mathbb{R}^{13}$ represents 13 timestamps within a 2-second window, with the current state positioned at the center of the window. Each phase value at timestamp $i$, channel $n$ is computed by 
\begin{equation}
    \mathrm{p}_{2i-1}=\mathrm{A}_i \cdot \sin \left(2 \pi \cdot \mathrm{S}_i\right), \quad \mathrm{p}_{2i}=\mathrm{A}_i \cdot \cos \left(2 \pi \cdot \mathrm{S}_i\right)
\end{equation}
where $\mathrm{A}_i, \mathrm{S}_i$ is the learned latent parameters. The motion style is embedded into $C^{\mathcal{S}} \in \mathbb{R}^{N \times 256}$ where $N$ is the number of the motion classes and $256$ is the dimensions encoded by an embedding layer. The embedding layer is a lookup table, encoding each distinct motion style from the dataset as a vector with a fixed dimension space based on the labeling categories. Finally a time embedding feature $ C^{\mathcal{T}}$ with dimension $d$ is constructed using the positional encoding as the following:
\begin{equation}
\mathbf{z}_{tta, 2i}=\sin \left(\frac{tta}{10000^{2i/d}}\right), \quad
\mathbf{z}_{tta, 2i+1}=\cos \left(\frac{tta}{10000^{2i/d}}\right)
\end{equation}
where $i \in [0,...,d/2]$ and $d$ is the time-to-arrive embedding dimension which we set as $128$ in our implementation. 
The gating network is a three-layer fully-connected network with [512, 128, 16] units followed by SiLU activation, where the output dimension $16$ is equal to the number of experts. Then the output coefficients computed by the gating network are blended with the outputs of expert networks linearly to get the pose feature in the next frame.

\subsection{Losses}
To obtain realistic results, we train our DC-MoE with multi-task losses, including reconstruction loss and kinematic consistency loss. 

\paragraph{Reconstruction Loss.} The reconstruction loss between the predicted next frame pose and ground truth from the motion clip $L_{recon}$ is defined using the $L2$ norm, where $\hat{\mathbf{p}}_n, \hat{\mathbf{r}}_n, \hat{\mathbf{v}}_n $ represent the model's estimated joint position, joint rotation, and joint linear velocity in the next frame. $\hat{\mathbf{c}}_t, \hat{\mathbf{t}}_t, \hat{\mathcal{P}}_t$ are the model's estimated joint contact, root trajectory, and phases in the time series window $T=7$ of the future $1$ second:
\begin{equation}
\begin{aligned}
& L_{\text{pos}} = \frac{1}{N} \sum_{n=1}^{N} \| \hat{\mathbf{p}}_n - \mathbf{p}_n \|_2   
& L_{\text{rot}} = \frac{1}{N} \sum_{n=1}^{N} \| \hat{\mathbf{r}}_n - \mathbf{r}_n \|_2  \\
& L_{\text{vel}} = \frac{1}{N} \sum_{n=1}^{N} \| \hat{\mathbf{v}}_n - \mathbf{v}_n \|_2   
& L_{\text{contacts}} = \frac{1}{T} \sum_{t=1}^{T} \| \hat{\mathbf{c}}_t - \mathbf{c}_t \|_2 \\
& L_{\text{traj}} = \frac{1}{T} \sum_{t=1}^{T} \| \hat{\mathbf{t}}_t - \mathbf{t}_t \|_2 
& L_{\text{phase}} = \frac{1}{T} \sum_{t=1}^{T} \| \hat{\mathcal{P}}_t - \mathcal{P}_t \|_2 
\end{aligned}
\end{equation}

\paragraph{Kinematic Consistency Loss.} To strengthen the physical realism of the prediction pose, we adopt kinematic consistency loss $L_{consist}$ similar to those in \cite{ICLR2023:MDM,cvpr2023:EDGE} as auxiliary losses:

\begin{align}
L_{\text{contacts}^{'}} &= \frac{1}{N} \sum_{N=1}^{N} \| \hat{\mathbf{c}}_n \cdot \hat{\mathbf{v}}_n \|_2 \label{eq:contact_consistant}\\
L_{\text{pos}^{'}} &= \frac{1}{N} \sum_{n=1}^{N} \| (x_{n} + \hat{\mathbf{v}}_n  \cdot \Delta t) - \hat{\mathbf{p}}_n \|_2^{2} \label{eq:pose_consistant}
\end{align}

In Eq.\eqref{eq:contact_consistant} we calculate a contact consistency loss similar to \cite{cvpr2023:EDGE}. Here the model's own binary prediction of foot contact $\hat{\mathbf{c}}_n$ penalized the foot velocity where the prediction shows foot sliding is invalid. We also add another velocity consistency term Eq.\eqref{eq:pose_consistant}, where the estimated linear velocity is applied to the current input pose $x_n$ and the computed position is compared with the model's estimate $\hat{\mathbf{p}}_n$ in the next frame.

The overall training loss for DC-MoE is the weighted sum of the reconstruction and kinematic consistency loss:
\begin{equation}
    L = \lambda_{recon} L_{\text{recon}} + \lambda_{consist} L_{\text{consist}}
\end{equation}

\section{Trajectory Guidance}
In this section, we describe the definition of the trajectory gallery used for in-betweening motion synthesis. Then we give the scenarios for trajectory guidance based on the trajectory gallery, which we refer to as an arbitrary path search with varied durations.

\subsection{Trajectory gallery}
For planning the global translation, a trajectory gallery is constructed. We can define a matching gallery $\mathbb{G}$ similar to Learned Motion Matching\cite{sig20:learnedmotionmatching} but in a simplified form with reduced dimensions. This can be achieved by traversing the entire database of the root trajectory utilizing a sliding window. The size of the sliding window ranges from 15 to 150 frames, corresponding to the duration between 0.5s to 5s. Each of these extracted trajectories represents a fundamental unit of movement starting from the origin world coordinate, termed as an "atomic trajectory". We define an atomic trajectory as a tuple $ \mathit{T}=\{id, {o^{s}, o^{e},v^{p}, c}\} \in \mathbb{R}^{9}\ $, {$id= (t_s, t_e) $} is the start frame and end frame index in the database respectively, $o^{s} \in \mathbb{R}^{2}$ is the unit vector indicating the relative root facing orientation projected on the ground, $o^{e} \in \mathbb{R}^{2}$ is the unit vector indicating the relative root ending orientation projected on the ground, $v^{p} \in \mathbb{R}^{2}$ is the vector pointing from the start position to the end position, and $c \in \mathbb{R}$ is the additional condition for style, such as jogging, dancing, and crawling.

\subsection{Arbitrary trajectory matching}
The goal of arbitrary trajectory matching is to find a set of trajectories ($\geq 1$ atomic trajectory) corresponding to a specific search query. We define the search query as a tuple $\mathrm{q} = \{o^{s}, o^{e}, v^{p}\} \in \mathbb{R}^{6}$, where $o^{s}$ and $o^{e}$ are the beginning and end authoring pose root relative orientations projected on the ground. $v^{p}$ is the vector pointing from the beginning authoring pose to the end authoring pose. The matching mechanism aims to reach the given target poses while minimizing differences in the root-facing orientation within the error threshold.
We define the error as follows: \\
\begin{equation}
\begin{aligned}
Error\biggl((\mathit{T}_1, ..., \mathit{T}_n), \mathrm{q}\biggl) &= \text{angle}\Bigl(o^s(\mathit{T}_1), o^s(\mathrm{q})\Bigr) \\
&+ \text{angle}\Bigl(o^e(\mathit{T}_n), o^e(\mathrm{q})\Bigr)
\end{aligned}
\end{equation}
where $(\mathit{T}_1, ..., \mathit{T}_n)$ are the candidate trajectories and $\mathrm{q}$ is the searching query for two adjacent authoring frames. The $k^\text{th}$ best fit candidate for searching query $\mathrm{q}$ is defined as follows:
\begin{equation}
\hat{\mathit{T}_k} = {\arg \min}_{\mathit{T}\in \mathbb{G}\setminus \{\hat{\mathit{T}_1}, ..., \hat{\mathit{T}_{k-1}}\}} Error(\mathit{T}, \mathit{T}_n)
\end{equation}

For details of the combined optimization of the trajectory candidates please see Algorithm 1. 
Given a single query (a pair of two authoring poses), the trajectory matching proceeds the searching within the trajectory gallery using two manners. The initial search is first performed in batches on the gallery that have the same beginning-to-end distance as the query. Any candidates that meet the error threshold in this initial search are termed direct candidates.
When no direct candidates are found, the beginning-to-end distance in the query gets split into smaller sub-distances. This splitting employs a random sample from the trajectory distribution. Following this, the search is then executed recursively.

We find that 5 rounds of searching are already sufficient to get a composed trajectory within the error threshold when the authoring pose distances range from $0.1 \sim 10$ meters. To further enhance interaction during runtime, we use the K-means algorithm to categorize all the candidates into $3$ distinct groups by minimizing intra-cluster variance. Each group is determined based on the number of trajectory frames for each candidate. Post categorization, based on the characteristic nature of the groups, assigns corresponding time labels: "fast", "medium", and "slow" duration. This allows the user to assign the speed conditions manually using a semantic label.

\begin{algorithm}
% Algorithm
\begin{algorithmic}[1]
\State \textbf{input} $M$: database of the trajectory
\State \textbf{input} $q$: query for searching
\State \textbf{input} $\alpha$: threshold for total error
\State \textbf{input} $k$: number of keeping branches  
\State \textbf{output} $\mathit{T} = (\mathit{T}_1, ..., \mathit{T}_n)$: returning trace %, return empty trace if does not find one 
\State \textbf{output} $E: \text{Error}\Bigl((\mathit{T}_1, ..., \mathit{T}_n), q \Bigr)$
\State $DirectCandidates \gets$ initialize empty direct candidate list 
\State $Candidates \gets$ initialize empty candidate list 

\item[]
% This can run in parallel
\For {trajectory $\mathrm{y}$ in $M = \{\mathrm{y}_1, ..., \mathrm{y}_m\}$} 
    \If{$||v^p(y)|| = ||v^p(q)||$}    
        \State $\mathrm{y} \gets$ Rotate($\mathrm{y}, v^p(q)$)
        \State $E \gets \text{Error}(\mathrm{y}, q)$
        \If{$E \leq \alpha$}
            \State $DirectCandidates \gets Append(\mathrm{y}, E)$ 
            % \State $\mathit{T} \gets \mathrm{y}$
            % \State \Return
        \EndIf
    \ElsIf{$||v^p(y)|| < ||v^p(q)||$}
        \State $\mathrm{y} \gets$ Rotate($\mathrm{y}, v^p(q)$)
        \State $E' \gets \text{StartingError}(\mathrm{y}, q)$ 
        \If{$E' \leq \alpha$}
            \State $Candidates \gets Append(\mathrm{y}, E')$ 
        \EndIf
    \EndIf
\EndFor
\If{$len(DirectCandidates) > 0$} 
    \State $\mathit{T}, E \gets$ Sample Candidate($DirectCandidates$) %\Comment{random sample here}
    \State \Return
\EndIf
\item[]
\State Filtrate Candidates($Candidates, k$) \Comment{random sample sub-distance}
\For {pair $(\mathrm{y}, E')$ in $Candidates$}
    \State $\mathit{T}, E \gets$ TCS($M, \text{Subtract($q, \mathrm{y}$)}, \alpha-E', k$)
    \If{$len(\mathit{T}) > 0$}
        \State $\mathit{T} \gets (\mathrm{y}, \mathit{T})$
        \State $E \gets E + E'$
        \State \Return
    \EndIf
\EndFor
\State $\mathit{T} \gets$ empty
\State \Return
\caption{Trajectory Candidate Search (TCS)}\label{alg:cap}
\end{algorithmic}
\end{algorithm}
\label{subsec:arbitary_route_planning}

\begin{figure*}[htb] % Use !t to place the figure at the top of the page
    \centering
    \includegraphics[width=\linewidth]{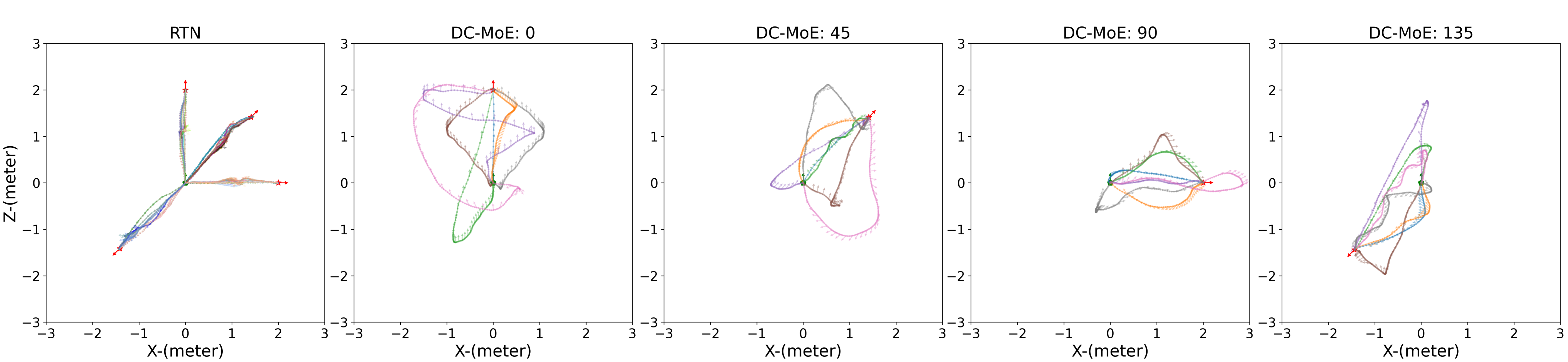} % Adjust width as needed
    \caption{Trajectory of the character's root produced by arbitrary route planning with different root orientations of the start and target pose under same start-to-end distance. The exploration in the RTN latent space does not produce varied trajectories and tends to collapse into an identical outcome. By utilizing random sampling from the trajectory gallery, our method allows the character to explore a wider range of in-between trajectories and diverse time durations. 
    \label{fig:experiment_diverse_traj}}
\end{figure*}

\section{Implementation}

\subsection{Training}
% Dataset, optimizer, Loss, and other hyper-param settings.
We use the LaFAN1\cite{Sig2020:RobustIbt} and 100Style  datasets\cite{2022:100StyleDataset} for training the DC-MoE, and we retarget the 100Style skeleton to LaFAN1 with the same 22 joints by removing one additional spine joint. A random sequence of target poses up to 3 seconds from the current frame is selected as the training pair (The DC-MoE framework is still capable of making extremely long transitions with good quality, more information is available in the supplementary material). We use the pre-trained periodic autoencoder to extract the phase feature of each frame in the dataset. For the LaFAN1 dataset, We split the dataset similarly to \cite{Sig2020:RobustIbt}
and \cite{SCA2023:PhaseInbetween}, where Subject 1 to 4 are used for training and Subject 5 for testing. The whole training set is augmented by mirroring all the motion. 
We train our network using the AdamW \cite{adamw:optimizer} optimizer with a learning rate of $1 e-4$, weight decay of $1 e-4$, and batch size of 32. The relative value for $\lambda_{recon}$ and $\lambda_{consist}$ in the loss term are set to 1 and 2.5 respectively. In order to simultaneously achieve target position reaching, we randomly mask the target bone position and rotation during the training. 

To incorporate the motion style condition with the training data, we distillate the LaFAN1 dataset into a sub-data in which motion styles are accurately described. Although the original LaFan1 dataset is prepared with labels such as "aiming", "dancing", and "fight" for the entire motion sequence, We find that there are some mismatched styles and non-style periods within the sequence. We leverage the method of Confident Learning\cite{JAIR:confident_learning}, which is a weak-supervised method based on the principle of noisy data pruning, to filter unsuitable motion sequences. We get a distilled dataset that contains $75.14\%$ samples of the full LaFan1 dataset. 

\subsection{Runtime}
We build our motion authoring system upon the open-source platform, derived from \cite{SCA2023:PhaseInbetween}. We deploy the trained model utilizing the external library ONNX to facilitate real-time inference. Our authoring system can run smoothly at over 40 frames per second on an RTX 3070ti GPU. 

\section{Experiments and Evaluation}

\subsection{Quantitative Comparisons}
We compare our work with the naive baseline interpolation, RTN \cite{Sig2020:RobustIbt} and Phase-IBT \cite{SCA2023:PhaseInbetween}, and report the evaluation metrics on the LaFan1 dataset with L2P and L2Q. The L2P and L2Q metrics respectively calculate the average $L2$ distance of each frame's joint global position and joint rotation in comparison to the ground truth. The naive interpolation uses linear interpolation for the root joint's position and spherically interpolation for all joints' rotation. These methods are then compared with our method. The experiment results can be found in Table \ref{tab:L2P_table1}. 

Compared with the baselines and previous methods (RTN and Phase-IBT), our proposed method achieves an overall better reconstruction accuracy. This is primarily due to the injected conditions that enhance the predictions' proximity to the ground-truth data, preventing the generated motions from being averaged. Additionally, we observe that RTN tends to drift towards the target without updating the joints when the in-between distance increases. When given stylized target poses, both RTN and Phase-IBT will fill the in-between frames with neutral locomotion poses such as running and walking. 

\begin{table}
    \centering
    \caption{benchmark results on LAFAN1 dataset }
    \begin{tabular}{c|ccc|ccc}
    \hline
    \multicolumn{7}{c}{L2P} \\
  \hline
    \multicolumn{1}{c}{Frames} &\multicolumn{1}{c}{15}  &\multicolumn{1}{c}{30} &\multicolumn{1}{c}{45} &\multicolumn{1}{c}{60} &\multicolumn{1}{c}{75} &\multicolumn{1}{c}{90}\\

       \hline
         Interp.  &3.32  &4.49  &5.46  &6.69  &8.09  &9.01  \\
         RTN      &\textbf{3.13}  &4.39  &4.82  &5.81  &6.07  &7.45  \\
         Phase-IBT&3.82  &4.26  &4.57  &5.42  &5.87  &6.55  \\
         Ours     &3.74  &\textbf{3.91}  &\textbf{4.41}  &\textbf{5.37}  &\textbf{5.64}  &\textbf{6.32}  \\

    \multicolumn{7}{c}{L2Q}\\
    \hline
    \multicolumn{1}{c}{Frames} &\multicolumn{1}{c}{15}  &\multicolumn{1}{c}{30} &\multicolumn{1}{c}{45} &\multicolumn{1}{c}{60} &\multicolumn{1}{c}{75} &\multicolumn{1}{c}{90}\\
        \hline
         Interp.  &1.52  &1.52  &1.67  &1.95  &2.23  &2.38  \\
         RTN      &1.09  &1.11  &1.12  &1.2   &1.75  &1.74  \\
         Phase-IBT&0.55  &0.59  &0.65  &0.78  &0.8   &0.88  \\
         Ours     &\textbf{0.46}  &\textbf{0.52}  &\textbf{0.57}  &\textbf{0.71}  &\textbf{0.75}  &\textbf{0.77}  \\
 \hline
    \end{tabular}
    \label{tab:L2P_table1}
\end{table}

\begin{table}[H]
    \centering
    \caption{Foot sliding metrics}
    \begin{tabular}{c|ccc|ccc}
    \toprule
         \multicolumn{7}{c}{Foot sliding}\\
         \cline{1-7}
         Frames&  15&  30&  45&  60&  75&  90\\
         Interp.&2.72  &3.19  &3.47  &3.84  &3.9  &4.47  \\
         RTN&0.61  &0.62  &0.66  &0.75  &0.91  &1.08  \\
         Phase-IBT&0.54  &0.58  &\textbf{0.6}  &0.67  &0.68  &0.71  \\
         Ours&\textbf{0.53} &\textbf{0.57}  &0.62  &\textbf{0.64}  &\textbf{0.64}  &\textbf{0.67}  \\
    \bottomrule
    \end{tabular}
    \label{tab:footSkate_table2}
\end{table}

We also analyze and compare the foot skating artifacts used in \cite{Sig2018:MANN_mix_Expert,SCA2023:PhaseInbetween}. Here, foot velocities, with the magnitude $v$ on the horizontal plane, are proportionately weighted by a contact-aware factor $s = v(2 - 2\frac{h}{H})$. This weighting is applied when the foot height $h$ falls within the maximum threshold of $H = 2.5\,cm$, to estimate the occurrence of skating in the motion sequences. The experiment results can be found in Table \ref{tab:footSkate_table2}.

\subsection{Trajectory gallery Analysis}
\begin{figure}[htb]
\includegraphics[width= \linewidth]{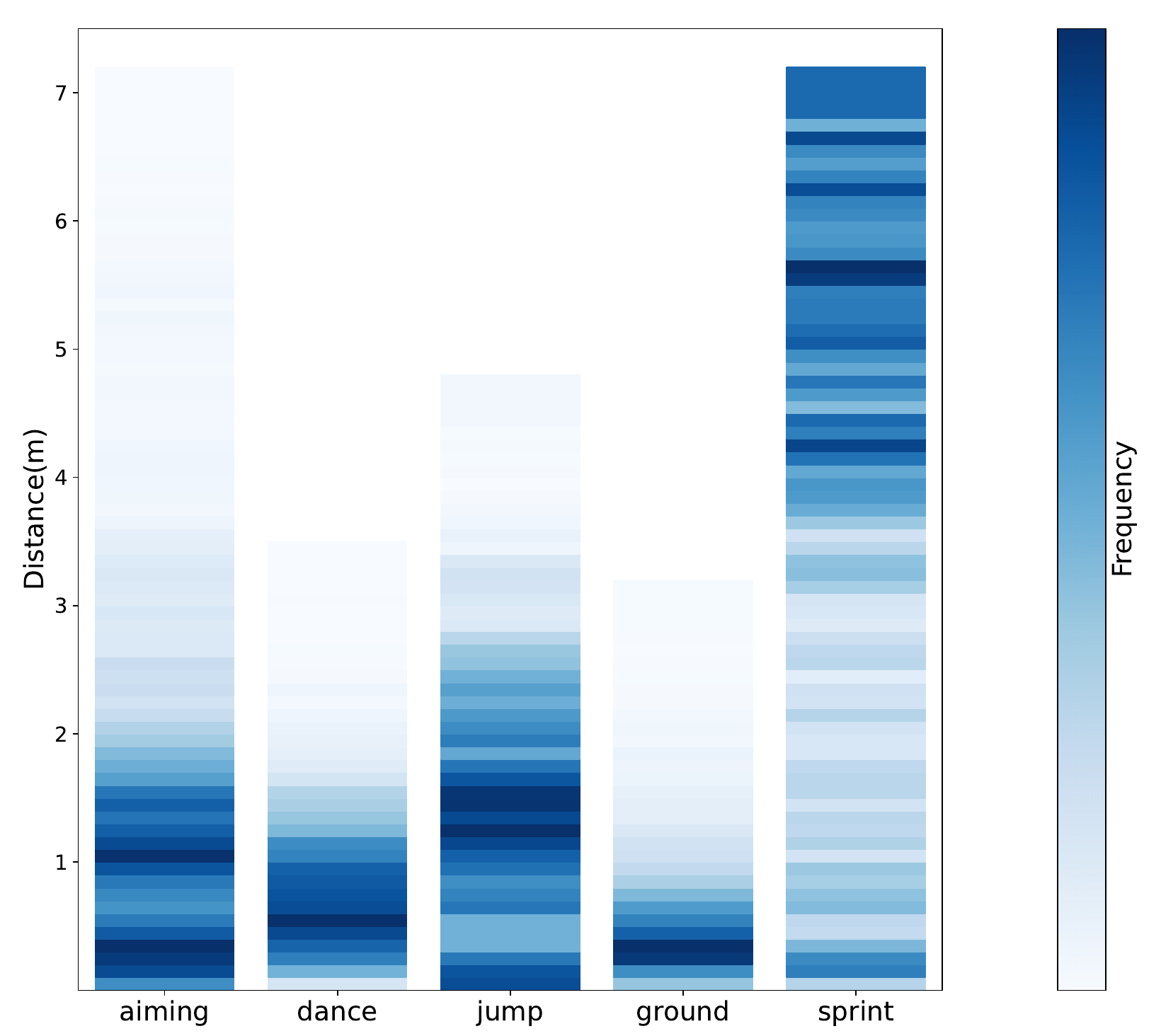}
\caption{Visualizing the distance distribution across different styles. Given the $45$ frames segments, we show the cumulative trajectory within this segment. Lighter colors denote fewer instances of a particular motion style, while darker colors indicate more numerous instances, which show the distribution of different motion styles. 
\label{fig: experiment_45frames_action_duration_distribution}}
\centering
\end{figure}

We further investigate the property of the trajectory gallery by setting the same trajectory duration, shown in Fig ~\ref{fig: experiment_45frames_action_duration_distribution}. One fundamental observation is that  different motions possess varied speed distributions. 
The speed difference across various movements can be depicted by the accumulated distance covered within the same duration. 
Viewed from another perspective, to standardize the distance between the initial and target pose, certain motions should be executed at a quicker pace while others should be played at a slower rate. This necessitates the inclusion of trajectory guidance, as the trajectory provides details about the number of time steps required to reach the specified target.

\subsection{Trajectory impact}
To evaluate the effects of the arbitrary trajectory matching, we visualize the runtime in-between results produced by our DC-MoE framework. Figure~\ref{fig:experiment_diverse_traj} illustrates the real trajectories generated by DC-MoE under the guidance of the trajectory matching algorithm, as described in Section \ref{subsec:arbitary_route_planning}. We set the facing orientation difference between the initial pose and the target pose as $0^{\circ}, 45^{\circ}, 90^{\circ},135^{\circ}$ respectively. For each pair of poses, we sample $7$ possible trajectories that satisfy the condition within varied time durations. 

Next, we compare the root trajectory generated by the RTN using the same pair of poses. To fairly compare the trajectory diversity, we follow the original re-sampling method\cite{Sig2020:RobustIbt} that utilizes the scheduled target noise for re-sampling the same transition. The in-betweening frames are set as 30, 40, 50, 60, 70,80 and 90, respectively. Figure~\ref{fig:experiment_diverse_traj} shows that enabling scheduled target noise in the latent space of RTN has a limited impact on the transition diversity, compared with our method.

% Head 2
\subsection{Ablation Study}
\begin{figure}[htb]
\includegraphics[width= \linewidth]{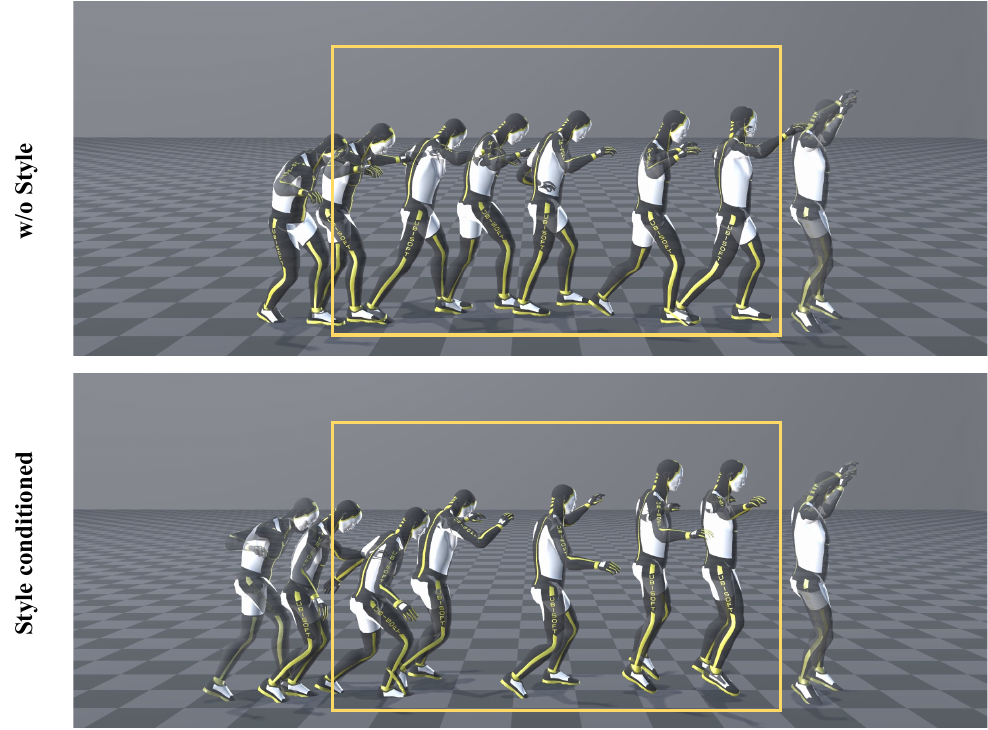}
\caption{ Example of with and without using style condition, while all other input features are kept same. The results indicate that with a style condition, the model can generate more dynamic motion.
\label{fig: style_impact}}
\centering
\end{figure}

\paragraph{Style condition} In Fig \ref{fig: style_impact}, a style condition case is shown, where the same set of initial pose and target pose is given. We visually compare the results generated with and without the style condition. Though both methods can produce natural stylized motion, by using style conditions, the model can produce a full-body movement more dynamically.  

\section{Limitations and Further work}
Like other data-driven approaches to kinematic motion synthesis, our in-between model is designed to predict poses that resemble those within the training set. However, when faced with unseen input key poses that present complex and unknown movements such as a backflip, the reasonable prediction of in-between remains a significant challenge.

Another promising research direction for in-between lies in the interaction with the environment. We notice that datasets like Lafan1 include the terrain changes, illustrated through actions like ascending stairs or jumping over boxes. Recent in-betweening motions do not take these environment interactions and constraints into consideration. It is interesting to explore how to equip kinematic motion synthesis systems with environment interaction abilities.

%Head1
\section{Conclusion}
In summary,  we have introduced a learning-based framework for data-driven in-between motion synthesis that is capable of autoregressively generating in-between motion from a single frame. By utilizing dynamic conditions, our framework can produce various high-fidelity motions under varied styles and durations. Our method can synthesize in-between motion on the fly, which enables real-time interaction in games and faster authoring for animators.
Moreover, the incorporation of explicit trajectory guidance empowers our framework to be more controllable in the process of in-between motion synthesis.

\begin{acks}
We would like to thank KangKang Yin, Xue Bin Peng, and Yi Shi for their various discussions and help.
\end{acks}

% \begin{acks}
% \end{acks}

\bibliographystyle{ACM-Reference-Format}
\bibliography{sample-base}

\clearpage
% \appendix

\section{Appendix}

\subsection{Arbitrary trajectory matching diagram}

\begin{figure}[H]
\centering
\includegraphics[width=\linewidth]{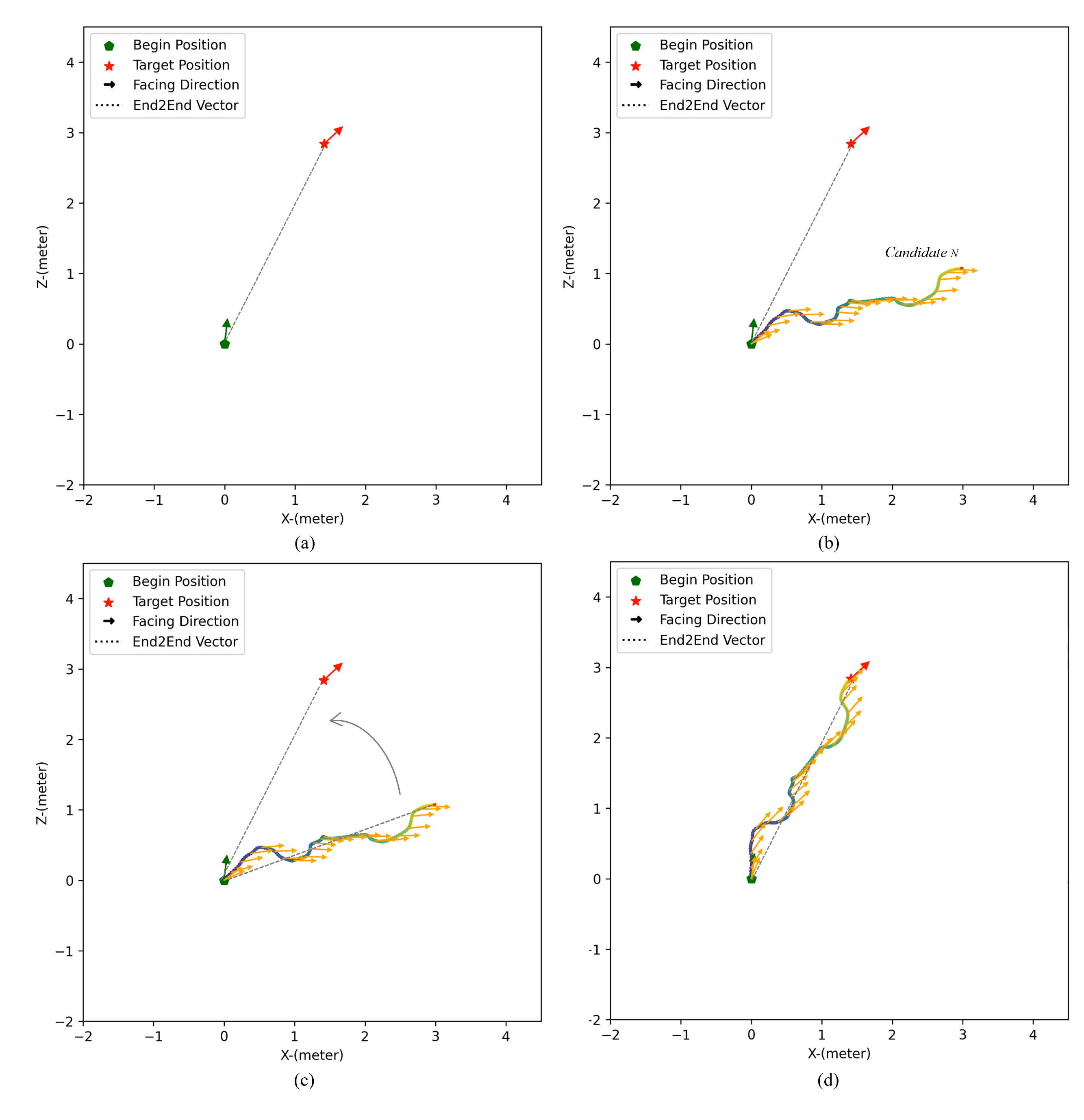}
\caption{Example of matching the candidate trajectory from the gallery with the same start-to-end distance as the input query.}
\label{fig:matching_diagram}
\end{figure}

We showcase a diagram of trajectory matching in Fig \ref{fig:matching_diagram}. For a given input query (a), which includes both the starting and target pose positions and orientations. In this instance, a candidate (b) with the same end-to-end distance as the query is selected. 
We then compute the angle between the two end-to-end vectors (c) to align them. The rotation matrix, which corresponds to this angle, is applied to each position and orientation on the candidate trajectory (d). In the process, we note the difference in start and end orientations between the candidate and the query.

\end{document}